\documentclass[twoside]{article}
\usepackage{qic,epsfig}
\usepackage{dcolumn}
\usepackage{bm}
\usepackage{amsfonts}
\usepackage{color}
\input xy
\xyoption{all}

\newcommand{\ket}[1]{\left| #1 \right\rangle}

\newcommand{\Id}{{\rm Id}}
\newcommand{\Stab}{\rm Stab}
\newcommand{\g}{\mathfrak{g}}
\newcommand{\gr}{\mathfrak{r}}

\begin{document}
\setlength{\textheight}{8.0truein}    

\runninghead{Symmetric states: local unitary equivalence via stabilizers}
            {Cenci, Lyons, Snyder, and Walck}

\normalsize\textlineskip
\thispagestyle{empty}
\setcounter{page}{1}


\vspace*{0.88truein}

\alphfootnote

\fpage{1}

\centerline{\bf
SYMMETRIC STATES: LOCAL UNITARY EQUIVALENCE VIA STABILIZERS}
\centerline{\footnotesize 
Curt D. Cenci}
\vspace*{0.015truein}
\centerline{\footnotesize\it Mathematics, Lebanon Valley College}
\baselineskip=10pt
\centerline{\footnotesize\it Annville, Pennsylvania 17003, USA}
\vspace*{10pt}
\centerline{\footnotesize
David W. Lyons}
\vspace*{0.015truein}
\centerline{\footnotesize\it Mathematics, Lebanon Valley College}
\baselineskip=10pt
\centerline{\footnotesize\it Annville, Pennsylvania 17003, USA}
\vspace*{10pt}

\centerline{\footnotesize 
Laura M. Snyder}
\vspace*{0.015truein}
\centerline{\footnotesize\it Physics, Lebanon Valley College}
\baselineskip=10pt
\centerline{\footnotesize\it Annville, Pennsylvania 17003, USA}
\vspace*{10pt}
\centerline{\footnotesize 
Scott N. Walck}
\vspace*{0.015truein}
\centerline{\footnotesize\it Physics, Lebanon Valley College}
\baselineskip=10pt
\centerline{\footnotesize\it Annville, Pennsylvania 17003, USA}
\vspace*{10pt}

\publisher{(received date)}{(revised date)}

\vspace*{0.21truein}

\abstracts{
We classify local unitary equivalence classes of symmetric states via a
classification of their local unitary stabilizer subgroups.  For states
whose local unitary stabilizer groups have a positive number of
continuous degrees of freedom, the classification is exhaustive.  We
show that local unitary stabilizer groups with no continuous degrees of
freedom are isomorphic to finite subgroups of the rotation group
$SO(3)$, and give examples of states with discrete stabilizers.  }{}{}

\vspace*{10pt}

\vspace*{3pt}
\communicate{to be filled by the Editorial}

\vspace*{1pt}\textlineskip    
\section{Introduction} 
 
The problem of measuring and classifying entanglement for states of
$n$-qubit systems is motivated by potential applications in quantum
computation and communication that utilize entanglement as a
resource~\cite{nielsen}.  On a deeper level, the phenomenon of
entanglement has played a key role in fundamental questions about
quantum mechanics itself.  Because entanglement properties of
multi-qubit states are invariant under local unitary (LU) transformations,
the entanglement classification problem leads naturally to the problem
of classifying local unitary equivalence classes of states.\\

In this paper we exploit the local unitary stabilizer group, that is,
the set of all LU transformations that fix a given state, as
a tool for local unitary equivalence classification.  
Given states $\ket{\psi}, \ket{\psi'}$ that are local equivalent via
some LU transformation $U$, say $\ket{\psi'}=U\ket{\psi}$,
their stabilizers are related as follows.  If $V\ket{\psi}=\ket{\psi}$,
then $UVU^\dagger\ket{\psi'}=\ket{\psi'}$.  Thus the stabilizer groups
for the two states are isomorphic via conjugation by $U$.  Thus the
isomorphism class of the stabilizer is an LU invariant.\\

Local unitary equivalence classification is a hard problem.  The number
of parameters needed to distinguish inequivalent states grows
exponentially with the number of qubits.  However it is possible to
obtain useful classification results for certain classes of states.
Recently, symmetric states (states that are unchanged by any permutation
of qubits) have been a case of particular interest \cite{Aulbach,
  Martin, Assalini, Markham, Mathonet, Campbell, Prevedel,
  Wieczorek,Bastin}.  Bastin et al.~\cite{Bastin} have achieved a
stochastic local operations and classical communication (SLOCC)
classification for symmetric states by analyzing the Majorana
representation of symmetric states as symmetrized products of 1-qubit
states.  The SLOCC group contains the LU group, and therefore has fewer
equivalence classes.  In this paper, we present an LU classification
(partial in some cases) for symmetric states based on an analysis of
stabilizer structure.\\

The main results of this paper, stated more precisely and proved in
Section~3 below, are the following.  Theorem~1 classifies $n$-qubit
symmetric states whose local unitary stabilizers have a positive number
of continuous degrees of freedom.  The stabilizer must be one of four
possible types, with dimension $n$, $n-1$, 3, or 1. For each of these
stabilizer types, we give a complete description of the local unitary
equivalence classes of states with that stabilizer type.  Theorem~2
treats the case of symmetric states with discrete local unitary
stabilizers.  We show that any discrete stabilizer is isomorphic to a
finite subgroup of $SO(3)$. We construct examples of states with various
finite stabilizers and specific finite order stabilizing elements. \\

The paper is organized as follows.  Section~2 establishes notation and
conventions, summarizes the authors' previous 
results~\cite{minorb1, lyonsandwalck, walckandlyons, lyonswalckblanda,lyonsandwalck2} 
on stabilizer
structure that are needed in the present paper, and states and proves two
Lemmas used in the proofs of the main results.  Section~3 is devoted to
the two main Theorems described in the previous paragraph.  In the
concluding Section~4, we point out directions for future
investigation.\\

\section{Preliminaries}

\subsection{Notation and conventions}

Let $G$ denote the local unitary group $G=U(1)\times (SU(2))^n$ which
acts on the space ${\cal H}=({\rm \textbf C}^2)^{\otimes n}$ of
$n$-qubit state vectors.  Let $LG=u(1) \oplus \bigoplus_{k=1}^n su(2)$
denote the local unitary Lie algebra, where $u(1)$ is the set of pure
imaginary complex numbers and $su(2)$ is the set of $2\times 2$
traceless skew-Hermitian matrices.

We shall write $g^{(k)}$ or $N^{(k)}$ to denote the element $(1,
\Id,\ldots,g,\ldots,\Id)$ of the local unitary group or the element
$(0,0,\ldots,N,\ldots,0)$ of the local unitary algebra with the single entry
$g$ in $SU(2)$ or $N$ in $su(2)$, respectively, in the $k$-th position
(we count the phase factor as the $0$-th position).  We define the
weight of an element $(e^{it},g_1,\ldots,g_n)$ in $G$ or
$(it,M_1,\ldots,M_n)$ in $LG$ to be the number of coordinates $k$ in
positions $1$ through $n$ for which $g_k\neq \Id$ or $M_k\neq 0$.

The actions of
$g^{(k)}$ and $N^{(k)}$ on a product $\ket{\psi_1}\ket{\psi_2}\cdots
\ket{\psi_n}$ of 1-qubit states are given by
\begin{eqnarray*}
g^{(k)} (\ket{\psi_1}\ket{\psi_2}\cdots \ket{\psi_n}) 
&=& \ket{\psi_1}\cdots (g\ket{\psi_k}) \cdots \ket{\psi_n}\\
N^{(k)} (\ket{\psi_1}\ket{\psi_2}\cdots \ket{\psi_n}) 
&=& \ket{\psi_1}\cdots (N\ket{\psi_k}) \cdots \ket{\psi_n}
\end{eqnarray*}
and these actions extend linearly to all of ${\cal H}$.
The actions of $U=(e^{it},g_1,\ldots,g_n)$ in $G$ and $M=(it,M_1,M_2,\ldots,M_n)$ in $LG$ on $\ket{\psi}$ in ${\cal H}$ are given by
\begin{eqnarray}\label{gpaction}
U\ket{\psi}
&=& \left(e^{it}\prod_{k=1}^n g_k^{(k)}\right)\ket{\psi}\\ \label{algaction}
M\ket{\psi}
&=& \left(it+\sum_{k=1}^n M_k^{(k)}\right)\ket{\psi}.
\end{eqnarray}

We write $\Stab_\psi$ and $K_\psi$ to denote the local unitary
stabilizer subgroup and the local unitary stabilizer subalgebra for the
state $\ket{\psi}$, respectively, given by
\begin{eqnarray*}
\Stab_\psi &=& \{g\in G\colon g\ket{\psi}=\ket{\psi}\}\\
K_\psi &=& \{M\in LG\colon M\ket{\psi}=0\}.
\end{eqnarray*}
Because $\Stab_{\psi'} = \Stab_\psi$ and $K_{\psi'} = K_\psi$ whenever
$\ket{\psi'}$ is a scalar multiple of $\ket{\psi}$, we ordinarily do not
bother to normalize state vectors.
We denote elements of the computational basis for ${\cal H}$ using
$n$-bit multi-indices $I=i_1i_2\ldots i_n$ with $i_k=0,1$ for $1\leq
k\leq n$.  We shall make use of the standard basis $A=iZ$, $B=iY$,
$C=iX$ for $su(2)$, where $X,Y,Z$ are the Pauli matrices.
We shall write $\ket{D_n^{(k)}}$ to denote the weight $k$ unnormalized Dicke state
$$\ket{D_n^{(k)}}= \sum_{\mbox{\rm \small wt}(I)=k} \ket{I} $$ where
$\mbox{\rm wt}(I)$ is the number of ones in the bit
string $I=i_1i_2\ldots i_n$.  Any $n$-qubit symmetric state $\ket{\psi}$
can be written uniquely as a sum of Dicke states.
\begin{equation}\label{dickeexpand}
\ket{\psi} = \sum_{k=0}^n d_k \ket{D_n^{(k)}}.
\end{equation}

\subsection{Previous results on stabilizer structure}\label{stabstructres}

This subsection summarizes results on stabilizer structure from the
authors' previous papers~\cite{minorb1, lyonsandwalck, walckandlyons,
  lyonswalckblanda,lyonsandwalck2}.

Let $\g_k$ denote the $k$-th $su(2)$ summand of $LG$.  Given an $n$-qubit
state $\ket{\psi}$, the set $\{1,2,\ldots,n\}$ of qubit labels
decomposes into a disjoint union
\begin{equation}\label{qubitsetdecomp}
\{1,2,\ldots,n\} = {\cal B}_1 \cup {\cal B}_2 \cup \cdots \cup {\cal
  B}_p \cup {\cal R}_1 \cup {\cal R}_0
\end{equation}
with the following properties.
Each set ${\cal B}_k$, called an $su(2)$ block of qubits, is a minimal
set of qubits with the property that the intersection $K_\psi \cap
\left(\bigoplus_{b\in {\cal B}_k} \g_b\right)$ is isomorphic to $su(2)$.
The set ${\cal R}_1$
consists of all the qubits $k$ for which the projection $\pi_k\colon
K_\psi \to \g_k$ has a 1-dimensional image, and the set ${\cal R}_0$
consists of all the qubits $k$ for which the projection $\pi_k\colon
K_\psi \to \g_k$ has a 0-dimensional image.  Corresponding to the 
decomposition~(\ref{qubitsetdecomp}) is the decomposition of $K_\psi$ as a direct sum
\begin{equation}\label{kpsidecomp}
K_\psi = \left(\bigoplus_{i=1}^p {\mathfrak m}_i\right) \oplus \gr_1 \oplus \gr_0.
\end{equation}
Each algebra ${\mathfrak m}_i= K_\psi \cap \left(\bigoplus_{b\in {\cal
    B}_i} \g_b\right)$, called an $su(2)$ block for $K_\psi$ (because it
is isomorphic to $su(2)$), is spanned by elements $U=\sum_{b\in
  B_i}u_{b}$, $V=\sum_{b\in B_i}v_{b}$, $W=\sum_{b\in B_i}w_{b}$, with
each triple $u_b,v_b,w_b$ forming a basis of $\g_{b}$.  The algebra
$\gr_1=K_\psi \cap \left(u(1)\oplus \bigoplus_{k\in {\cal
    R}_1}\g_k\right)$ has the property that $\pi_k\colon K_\psi \to
\g_k$ is 1-dimensional for all $k\in {\cal R}_1$, and the algebra
$\gr_0=K_\psi \cap \left(\bigoplus_{k\in {\cal R}_0}\g_k\right)=0$ has the
property that $\pi_k\colon K_\psi \to \g_k$ is 0-dimensional for all
$k\in {\cal R}_0$.\\

In the proof of the Theorem~\ref{stabpsiinfinite}, it will be necessary
to use an expression established in~\cite{lyonsandwalck2} for a
state $\ket{\psi}$ whose stabilizer is the ``standard'' $su(2)$ block
\begin{equation}\label{stdsu2block}
K_\psi=\{(0,M,M,\ldots,M)\colon M\in su(2)\}.
\end{equation}
In this case it is
necessarily true that the number $n$ of qubits is an even number $2m$.
Given a partition ${\cal P}$ of the set $\{1,2,\ldots,n\}$ of qubit
labels into $m$ subsets of size 2, we define the state $\ket{s_{\cal P}}$ to be the product of singlets
\begin{equation}\label{sp}
\ket{s_{\cal P}} = (\ket{01}-\ket{10})^{\otimes_{\cal P} m} 
\end{equation}
where the subscript ${\cal P}$ on the tensor symbol means that the 
  pairs of qubits that form the singlet factors of $\ket{s_{\cal P}}$
  are the elements of ${\cal P}$.  For example, for 
${\cal P}=\{\{1,4\},\{2,3\}\}$, the state $\ket{s_{\cal P}}$ is the tensor
  product of a singlet in qubits $1,4$ times a singlet in qubits $2,3$.
To each partition ${\cal P}$ we associate a diagram consisting of
$m$ chords on a circle, as follows.  Mark a circle with vertices labeled
$\{1,2,\ldots,n\}$ in order around the circle.  For each element
$\{i,j\}$ in ${\cal P}$, draw a chord connecting $i$ and $j$.  We say
${\cal P}$ {\em has no crossings} if none of the chords intersect in its
associated diagram.  For example, ${\cal P}=\{\{1,4\},\{2,3\}\}$ has no
crossings, but ${\cal P}'=\{\{1,3\},\{2,4\}\}$ has a crossing.  See
Figure~\ref{chorddiagrams}.
\newcommand{\pnocross}{$ {\cal P} = \{  \{1,4\} , \{2,3\} \} $}
\newcommand{\pcross}{$ {\cal P}' = \{  \{1,3\} , \{2,4\} \} $}
\newcommand{\spnocross}{$ \ket{s_{\cal P}} = \ket{0011} - \ket{0101} -
  \ket{1010} + \ket{1100} $}
\newcommand{\spcross}{$ \ket{s_{\cal P'}} = \ket{0011} - \ket{0110} -
  \ket{1001} + \ket{1100} $}
\begin{figure}
\begin{center}
\input{chorddiagramsmall.pstex_t}
\end{center}
\fcaption{\label{chorddiagrams}Chord diagrams without and with crossings.}
\end{figure}

We prove in~\cite{lyonsandwalck2} that any $\ket{\psi}$ with $K_\psi$ of the
form~(\ref{stdsu2block}) can be uniquely written as a linear combination
\begin{equation}\label{cdc}
\ket{\psi}= \sum_{{\cal P }\colon \mbox{ \small no crossings}} c_{\cal P} \ket{s_{\cal P}}
\end{equation}
where the sum is over all partitions ${\cal P}$ with no crossings.  We
will call such a state a {\em chord diagram construction}, or CDC.  For
each ${\cal P}$ with no crossings, we define $\ket{I_{\cal P}}$ to be
the first (in dictionary or binary order) standard computational basis
vector that occurs in the expansion 
\begin{equation}\label{spexpand}
\ket{s_{\cal P}} = \sum_I s_I \ket{I}
\end{equation}
of $\ket{s_{\cal P}}$ in the
computational basis.  We shall make use of the fact that if ${\cal P}'$
is another partition with no crossings and distinct from ${\cal P}$,
then $\ket{I_{\cal P}}$ does not appear in the expansion
of $\ket{s_{\cal P'}}$ in the standard basis.  
It is easy to see that
$\ket{I_{\cal P}}$ is given by
\begin{equation}\label{ipformula}
  \ket{I_{\cal P}} = \ket{01}^{\otimes_{\cal P} m}
\end{equation}
and also that the basis vector $\ket{I_{\cal P}^c}$ for the bitwise
complement of $I_{\cal P}$, given by
\begin{equation}\label{ipcompformula}
  \ket{I_{\cal P}^c} = \ket{10}^{\otimes_{\cal P} m}
\end{equation}
also has the property that it does not appear in the expansion of
$\ket{s_{\cal P'}}$ in the standard basis for all partitions ${\cal P}'$
with no crossings and distinct from ${\cal P}$.  Thus
we can write
\begin{equation}\label{ipexpand}
  \ket{\psi}= \sum_{{\cal P}} c_{\cal P} \ket{s_{\cal P}}
= \sum_{\cal P} c_{\cal P}\left(\ket{I_{\cal P}} + (-1)^m\ket{I_{\cal P}^c}\right) + \sum_J s_J \ket{J}
\end{equation}
where no $J$ is equal to $I_{\cal P}$ or $I_{\cal P}^c$ for any ${\cal
  P}$.  In~(\ref{ipexpand}) and in the following subsection, we assume
the restriction that all partitions ${\cal P}$ have no crossings.

\subsection{Two lemmas for the main results}

\begin{lemma}\label{cdclemma}
No CDC is a symmetric state.
\end{lemma}

\medskip

\proof{Suppose on the contrary that $\ket{\psi}$ is a both a CDC and a
  symmetric state.  For $n=2$, the
  only CDC is $\ket{01}-\ket{10}$, which is not symmetric.  Now suppose
  that $n>2$.
  Equations~(\ref{dickeexpand}) and~(\ref{cdc}) give us two ways to
  write $\ket{\psi}$, so we have
$$\ket{\psi}=\sum_{k=0}^{n} d_{k} \ket{D_{n}^{(k)} } =
  \sum_{\mathcal{P}} c_{\mathcal{P}} \ket{s_{\mathcal{P}}} . $$
  Since $\ket{\psi}$ is a CDC, we know that $n=2m$ for some $m$.  Since
  the weight of each computational basis element in the expansion of
  $\ket{s_{\mathcal{P}}}$ is $m$, we can conclude that $d_{k} = 0$ for
  all $k\neq m$, giving us
$$ \ket{\psi}=d_{m} \ket{D_{n}^{(m)}} =  \sum_{\mathcal{P}}
  c_{\mathcal{P}} \ket{s_{\mathcal{P}}} . $$ 
Utilizing $I_{\cal P}$ defined in the previous subsection, the sum on
the right can be expanded in the computational basis
$$\sum_{\mathcal{P}} c_{\mathcal{P}} \ket{s_{\mathcal{P}}} = 
\sum_{\cal P} c_{\cal P}\ket{I_{\cal P}} + \sum_J s_J \ket{J}$$
where no $J$ is equal to $I_{\cal P}$ for any ${\cal P}$.
Thus we have $c_{\mathcal{P}} = d_{m}$ for all $\mathcal{P}$.  Now we
will show that $\ket{D_{n}^{(m)} } \neq \sum_{\mathcal{P}}
\ket{s_{\mathcal{P}}}$, contradicting our assumptions. There are two
cases: $m$ odd and $m$ even.\\

Let $m$ be odd.  By~(\ref{ipexpand}), computational basis vectors 
$\ket{I_{\cal P}}$ and $\ket{I_{\cal P}^c}$ appear with opposite signs
in $\sum_{\mathcal{P}} \ket{s_{\mathcal{P}}}$, so $m$ can not be odd.

Now, let $m$ be even.  We will demonstrate that exactly two of all the
partitions $\mathcal{P}$ have $s_{\widehat{I}} \neq 0$ in the expansion
of $\ket{s_{\mathcal{P}}}$ in~(\ref{spexpand}) where
$$\ket{\widehat{I}} = \vert{\underbrace{00 \ldots 001}_{\mbox{\small
      $m$ qubits}} \;\underbrace{011 \ldots 11}_{\mbox{\small $m$ qubits}}}\rangle.$$
Thus, if that basis element appears in $\sum_{\mathcal{P}}
\ket{s_{\mathcal{P}} }$ exactly twice, its coefficient can only be 0 or
$\pm$2 rather than the necessary $+$1.  The two partitions which are
responsible for the basis element are $\mathcal{P}_{0}$ and
$\mathcal{P}_{1}$, shown in Fig.~\ref{partitions1}.

\begin{figure}
\begin{center}
\input{partitions1small.pstex_t}
\end{center}
\fcaption{\label{partitions1}Chord diagrams of partitions $\mathcal{P}_0$ and $\mathcal{P}_1$ respectively}
\end{figure}

Suppose we have a partition $\mathcal{P}$ with nonzero
$s_{\widehat{I}}$.  Referring to Fig.~\ref{partitions2}, we would need
to connect each open circle to a filled circle with a line segment while
avoiding all intersections in order to produce such a basis element. All
such configurations will yield the partitions which produce
$\ket{\widehat{I}}$.  We now wish to show that $\mathcal{P} =
\mathcal{P}_{0}$ or $\mathcal{P} = \mathcal{P}_{1}$.

\begin{figure}
\begin{center}
\input{partitions2.pstex_t}
\end{center}
\fcaption{\label{partitions2}Associate filled dots with 1's and open dots with 0's.}
\end{figure}

Consider the vertex at $m$.  We cannot connect this filled circle to any
of the open circles 1 through $m-2$ since we would effectively ``trap''
open circles with the added line segment.  Thus, $m$ clearly must
connect to either $m-1$ or $m+1$.  Consider first the case where the
vertices at $m$ and $m-1$ are connected.  This situation leaves $m+2$ as
the only viable option for connection with $m+1$ since a line segment
from $m+1$ to any of $m+3$ through $n$ would trap filled circles.  For
the same reason, we can deduce that 1 must connect with $n$ to avoid a
trap, and then 2 must connect with $n-1$, so on and so forth until $m-2$
must connect to $m+3$.  This configuration is $\mathcal{P}_{1}$ exactly.
Now we examine the implications of a connection between $m$ and $m+1$.
Just as before, 1 and $n$ are forced into a connection, then 2 with
$n-1$, $\ldots$ , and $m-1$ must connect to $m+2$.  This configuration
is $\mathcal{P}_{0}$ exactly.  Hence, if a partition $\mathcal{P}$ of
2-element subsets of the set $\{1, \; 2, \; \ldots \; , \; n \}$ has
$s_{\widehat{I}} \neq 0$, then $\mathcal{P}$ is either the
aforementioned $\mathcal{P}_{0}$ or $\mathcal{P}_{1}$.

Thus, we can conclude that no $n$-qubit CDC is symmetric for $n>2$.
}

\begin{lemma}\label{su2blocklemma}
Suppose $K_{\psi}$ has an $su(2)$ block ${\mathfrak m}$ in qubits ${\cal B}$.  
Then for every nonzero stabilizing element $M=\sum_{b\in {\cal B}}M_b^{(b)}$ we have
$M_{k} \neq 0$ for all $k\in {\cal B}$.
\end{lemma}

\medskip

\proof{Let $M$ in $K_\psi$ be given as above and let $U$, $V$, and $W$
  span ${\mathfrak m}$ as described in the previous subsection.  If
  $M=aU+bV+cW$, then $M_k = au_k + bv_k + cw_k$ for all $k$ in ${\cal B}$.  Therefore
  each $M_k$ is nonzero.}

\section{Main results}

The main results are consequences of the stabilizer
decomposition~(\ref{kpsidecomp}) together with the assumption that
$\ket{\psi}$ is symmetric.  Here is an outline.  Symmetry implies that
the stabilizer $K_\psi$ must consist of a single summand
in~(\ref{kpsidecomp}).  If that summand is a single $su(2)$ block
algebra ${\mathfrak m}$, then $\ket{\psi}$ must be a 2-qubit (LU
equivalent) singlet state.  If $K_\psi$ is the summand $\gr_1$, then
$\ket{\psi}$ is either a product state, a generalized GHZ LU equivalent,
or LU equivalent to a single Dicke state.  Finally, if $K_\psi$ is the
summand $\gr_0=0$, then $\Stab_\psi$ is a finite subgroup of $SO(3)$.  We
do not claim to have a full classification of LU equivalence classes of
states in this last category, but we give examples that suggest
interesting questions. 

The first subsection below treats the cases $K_\psi ={\mathfrak m}$ and
$K_\psi = \gr_1$.  This is equivalent to $K_\psi$ being a nonzero
dimensional algebra, which is in turn equivalent to $\Stab_\psi$ being a
nonzero dimensional Lie group, which is in turn equivalent to
$\Stab_\psi$ being an infinite set. The second subsection treats the
case $K_\psi=\gr_0=0$.  This is equivalent to $\Stab_\psi$ being a
finite group.  

\subsection{Continuous local unitary stabilizers}

\begin{theorem}\label{stabpsiinfinite}
Let $\ket{\psi}$ be an $n$-qubit symmetric state whose local unitary
stabilizer $\Stab_\psi$ is infinite.  Then one of four possibilities
holds.

\begin{enumerate}
\item[(i)] The dimension of $\Stab_\psi$ is $n$, the state $\ket{\psi}$
  is LU equivalent to the product state $\ket{00\cdots 0}$, and $K_\psi$
  is LU equivalent to
$$K_{\mbox{\small \rm product}}=\langle -i + A^{(1)},-i + A^{(2)},
  \ldots, -i + A^{(n)}\rangle.$$ There is one equivalence class of this
  type.

\item[(ii)] The dimension of $\Stab_\psi$ is $n-1$, the state
  $\ket{\psi}$ is LU equivalent to a generalized GHZ state of the form
  $a\ket{00\cdots 0}+b\ket{11\cdots 1}$ for some $n>2$, and $K_\psi$ is
  LU equivalent to
$$K_{\mbox{\small \rm GHZ}}=\langle A^{(1)} - A^{(2)},A^{(1)} - A^{(3)},
  \ldots, A^{(1)} - A^{(n)}\rangle.$$ 
We may take $a,b$ to both be positive
  and real with $a\geq b$.  The LU equivalence classes of this type are
  parameterized by the interval $0<t\leq 1$ by $a=\cos \frac{\pi}{4}t$,
  $b=\sin \frac{\pi}{4}t$.

\item[(iii)] The dimension of $\Stab_\psi$ is 3, the state $\ket{\psi}$
  is LU equivalent to the singlet state $\ket{01}+\ket{10}$, and
  $K_\psi$ is LU equivalent to
$$K_{\ket{01}+\ket{10}}=\langle A^{(1)}+A^{(2)}, B^{(1)}-B^{(2)},
  C^{(1)}-C^{(2)}\rangle.$$ There is one LU equivalence class of this
  type.

\item[(iv)] The dimension of $\Stab_\psi$ is 1, the state $\ket{\psi}$
  is LU equivalent to a single Dicke state $\ket{D_n^{(k)}}$ for some
  $n>2$ and some $k$ in the range $0< k< n$, and $K_\psi$ is LU
  equivalent to
$$K_{D_n^{(k)}}=\langle i(n-2k)+A^{(1)} + A^{(2)} + \cdots + A^{(n)}\rangle.$$
  There are $\lfloor n/2\rfloor$ LU equivalence classes of this type,
  with representatives
 $$\ket{D_n^{(1)}},\ket{D_n^{(2)}},\ldots,\ket{D_n^{(\lfloor{n/2}\rfloor)}}.$$

\end{enumerate}

\end{theorem}

{\em Comments:} The true $n$-qubit GHZ state with $|a|=|b|$ is distinguished among the generalized GHZ states with $|a|\neq |b|$ of type~(ii) above by additional discrete symmetry.  Its Pauli stabilizer group has size $2^n$ and is generated by weight two Pauli group elements of the form $Z^{(i)}Z^{(j)}$ together with the element $XX\cdots X$, while the Pauli stabilizer of the generalized GHZ with $|a|\neq |b|$ has size $2^{n-1}$ and is generated completely by the elements $Z^{(i)}Z^{(j)}$.  Similarly, the Dicke state $\ket{D_n^{(n/2)}}$ for $n$ even is distinguished from other Dicke states $\ket{D_n^{(k)}}$ of type~(iv) above by additional discrete symmetry.  The stabilizer $\Stab_{D_n^{(n/2)}}$ contains the elements
$$(-1)^{(n/2)} ZZ\cdots Z, XX\cdots X, YY\cdots Y$$
whereas for $k\neq n/2$, we have only the Pauli $Z$s, but not the $X$s and $Y$s. \\


\proof{Let $\ket{\psi}$ be an $n$-qubit symmetric state such that
  $Stab_{\psi}$ is infinite.  As discussed above, this is equivalent to
  $\Stab_\psi$ and $K_\psi$ having positive dimension.  

Our first claim is that the decomposition~(\ref{kpsidecomp}) for $K_\psi$
can have only one summand for symmetric $\ket{\psi}$.  
Permutation invariance implies that the dimension of the projection of
$K_\psi$ into any $\g_k$ factor of $LG$ must be the same for all qubits.
This rules out the possibility of having $\gr_0$ or $\gr_1$ with any
other summand in $K_\psi$, but there remains the possibility that
$K_\psi$ could be the sum of two or more $su(2)$ blocks.  If this were
the case, say $K_\psi$ has $su(2)$ blocks in qubits ${\cal B}_1$ and
${\cal B}_2$, then we would have a nonzero element in $K_\psi$ of the
form $it+\sum_{i=1}^{n} \chi_{i} M^{(i)}$ where $\chi_{i}$ is 1 when the
$i$-th qubit is in ${\cal B}_1$  and 0 otherwise.  Now, pick an
$i_{0}$ and $i_{1}$ where $i_{0}$ is in ${\cal B}_1$ and $i_{1}$ is not.
Since $\ket{\psi}$ is symmetric, we can transpose $\chi_{i_{0}}$ and
$\chi_{i_{1}}$ in $it + \sum_{i=1}^{n} \chi_{i} M^{(i)}$ and have
another stabilizing element.  However, this contradicts
Lemma~\ref{su2blocklemma}.  Therefore $K_{\psi}$ cannot contain more
than one $su(2)$ block.

Now suppose that $K_\psi$ is a single $su(2)$ block, and let
$M=(it,M_{1},M_2,\ldots,M_n)= it+\sum_k M_k^{(k)}$ be a nonzero element
in $K_{\psi}$.  By symmetry, the element $M'=it+M_j^{(i)} + M_i^{(j)} +
\sum_{k\neq i,j}M_k^{(k)}$, obtained by transposing coordinates $i,j$,
is also in $K_\psi$.  Therefore we have $M-M'=\left( M_{i} -
M_{j}\right)^{(i)} + \left( M_{j} - M_{i}\right)^{(j)}$ also in
$K_\psi$.  By Lemma~\ref{su2blocklemma}, it is impossible for $M_{i} -
M_{j}$ to be nonzero for $n>2$ (because if $M_i-M_j\neq 0$, then $M-M'$
would be a weight $2$ stabilizing element, whereas
Lemma~\ref{su2blocklemma} says that any nonzero stabilizing element must
have full weight $n>2$). Therefore, for $n>2$ we have $M_{i} =
M_{j}$ for all $i$ and $j$.  Therefore all elements in $K_{\psi}$ are of
the form $N^{(1)} + N^{(2)} + \ldots + N^{(n)}$, so $\ket{\psi}$ is a
CDC.  But this contradicts Lemma~\ref{cdclemma}.  Therefore we conclude
that the only symmetric states with $su(2)$ block stabilizers must have
$n=2$ qubits.  There is only one LU equivalence class of 2-qubit states
with $su(2)$ block stabilizer, namely, the singlet state class.  This
establishes~(iii) of Theorem~\ref{stabpsiinfinite}.

Now we suppose that $K_\psi=\gr_1$, so that the projection of $K_\psi$
into each qubit factor $\g_k$ is 1-dimensional.  If $K_\psi$ contains a
weight one element $it+M^{(k)}$ then the $k$-th qubit is unentangled
\cite{lyonsandwalck}.  Since $\ket{\psi}$ is symmetric, it then must be
the case that every qubit is unentangled and $\ket{\psi}$ is LU
equivalent to $\ket{00...0}$.  This
establishes~(i) of Theorem~\ref{stabpsiinfinite}.

Now suppose that $K_{\psi}$ contains no weight one elements.
Suppose there exists a nonzero element $M=(it,M_{1},M_2,\ldots,M_n)=
it+\sum_k M_k^{(k)}$ in $K_\psi$ such that $M_i\neq M_j$ for some $i,j$.
Once again, as above, we have the element $M'=it+M_j^{(i)} + M_i^{(j)} +
\sum_{k\neq i,j}M_k^{(k)}$, obtained by transposing coordinates $i,j$,
also in $K_\psi$.  Therefore we have $M-M'=\left( M_{i} -
M_{j}\right)^{(i)} + \left( M_{j} - M_{i}\right)^{(j)}$ also in
$K_\psi$.  Let $N=M_i-M_j$.  We have $n-1$ linearly independent weight
two elements $N^{(1)}-N^{(k)}$, $2\leq k\leq n$, so the dimension of
$K_\psi$ is at least $n-1$.  If the dimension of $K_{\psi}$ is greater
than $n-1$, then there is an element in $K_{\psi}$ of the form $M''=it+
\sum_{k=1}^{n} a_{k} N^{(k)}$ that does not lie in the span of the
elements $N^{(i)}-N^{(j)}$.  But then we would have the weight one
element $\left(\sum_{k=1}^n a_k\right)N^{(k)} = M''+\sum_{k=2}^{n} a_{i}
\left(N^{(1)}-N^{(k)} \right)$ in $K_{\psi}$, which contradicts our
assumption.  Thus, we conclude that $K_{\psi} = \langle N^{(1)}-N^{(i)}
\rangle$.  From \cite{lyonswalckblanda}, we can conclude that
$\ket{\psi}$ is LU equivalent to a generalized GHZ state
$a\ket{00\ldots0} + b\ket{11\ldots1}$.  This establishes~(ii) of
Theorem~\ref{stabpsiinfinite}.

The last case to consider is where $M_{i} = M_{j}$ for all $i,j$, for
all $M=(it,M_1,M_2,\ldots,M_n)$ in $K_\psi=\gr_1$.  Let
$M=(it,N,\ldots,N)$ be a nonzero element in $K_\psi$.  Choose $g\in
SU(2)$ to diagonalize $N$ so we have $gNg^\dagger = \alpha A$ for some
real scalar $\alpha$.  Then the state $\ket{\psi'}=g^{\otimes
  n}\ket{\psi}$ has 1-dimensional stabilizer 
$$K_{\psi'} = \langle it_0+ A^{(1)}+A^{(2)}+\cdots +A^{(n)}\rangle$$
for some real $t_0$.
Write $\ket{\psi'}$ in the Dicke basis 
$\ket{\psi'}=\sum d_{k} \ket{ D_{n}^{(k)}}.$ It is easy to check that 
$M= it_0+ A^{(1)}+A^{(2)}+\cdots +A^{(n)}$ kills computational basis
vectors of weight $k=\frac{n+t_0}{2}$, and multiplies others by nonzero
scalars. It follows that $\ket{\psi'}$ is a multiple of $\ket{
  D_{n}^{(\frac{n+t_0}{2})}}$. This establishes~(iv) of
Theorem~\ref{stabpsiinfinite}, and completes the proof of Theorem~\ref{stabpsiinfinite}.
} 

\subsection{Discrete local unitary stabilizers}
\begin{theorem}\label{stabfinite}
Let $\ket{\psi}$ be an $n$-qubit symmetric state whose local unitary
stabilizer $\Stab_\psi$ is finite.  Then $\Stab_\psi$ is isomorphic to
a finite subgroup of $SO(3)$ ($3\times 3$ real orthogonal matrices).
\end{theorem}

\medskip

\proof{
We claim that for all $U=(e^{i\theta},g_1,g_2,\ldots,g_n)$ in 
$\Stab_{\psi}$,	that for all $i,j$ we have $g_i=\pm g_j$.  Given the claim, we can absorb the signs into the phase factor and write $U=e^{i\phi}g^{\otimes n}$ for some
  $g\in SU(2)$.	 Though	$\phi$ and $g$ are not unique, the correspondence that sends $U$ to the equivalence class $[g]$ of $g$ in the projective group $PSU(2)=SU(2)/(\pm \Id)$ is a well-defined injective homomorphism $\Stab_\psi \to PSU(2)\approx SO(3)$, where the latter isomorphism is the standard identification of (projective equivalence classes of) $2\times 2$ unitary matrices with rotations of the Bloch sphere. 
  
  To establish the claim, suppose on the contrary that there is some
  $U=(e^{i\theta}, g_1,g_2,\ldots,g_n)$ in $\Stab_{\psi}$ and a pair
  $i,j$ of positions such that $g_i \neq \pm g_j$.  Since $\ket{\psi}$
  is symmetric, we can transpose the positions of $g_i,g_j$ to obtain a
  new element $U' = e^{i\theta}g_1^{(1)}\cdots g_j^{(i)}\cdots
  g_i^{(j)}\cdots g_n^{(n)}$ in $\Stab_\psi$. Let $V=(U')^\dagger$, and
  let $h=g_j^\dagger g_i$.  Then we have
$$VU = h^{(i)}(h^\dagger)^{(j)}$$
in $\Stab_\psi$.  Choose $k\in SU(2)$ to diagonalize $h$, so we	have
$$\widetilde{h}=khk^\dagger = \left[\begin{array}{cc} e^{it}&0\\ 0 & e^{-it}\end{array}\right].$$

Let $\ket{\widetilde {\psi}}=k^{\otimes n} \ket{\psi}$ so that we have
$\widetilde{h}^{(i)}(\widetilde{h}^\dagger)^{(j)}$ in
$\Stab_{\widetilde{\psi}}$.  If there is an $I$ such that $\ket{I}$
appears with nonzero coefficient in the expansion of
$\ket{\widetilde{\psi}}$ in the standard basis and $I$ has opposite
entries in its $i$-th and $j$-th positions, then we have $e^{2it}=1$,
but then we have $\widetilde{h}=\pm \Id$, which in turn implies $g_i=\pm
g_j$.  But this violates our assumption on the $g_k$'s.  We conclude
that the only nonzero terms that occur in the expansion of
$\ket{\widetilde{\psi}}$ are for multi-indices $I=00\cdots 0$ and
$I=11\cdots 1$.  But this means that $\ket{\psi}$ is a product state or
a generalized GHZ state, which violates our assumption that $\Stab_\psi$
is finite.  With this contradiction, we have proved the Claim, and the
proof of Theorem~\ref{stabfinite} is complete.  } 

\medskip

{\bf Examples.} Using the Majorana representation~\cite{Bastin} it
is easy to construct symmetric states with finite order stabilizer
elements.  The 3-qubit state
$$\ket{\psi}=\sum_\pi \pi\left[\ket{a}\ket{b}\ket{c}\right]$$
where the
sum ranges over all permutations $\pi$ of the set of qubits
and $\ket{a}$, $\ket{b}$, $\ket{c}$ are the 1-qubit states
$$
\ket{a}=\ket{0},\hspace*{.2in}\ket{b}=\ket{1},\hspace*{.2in}\ket{c}=\frac{1}{\sqrt{2}}(\ket{0}+\ket{1})
$$ is made from three points on the Bloch sphere that form vertices of
an isoceles, but not an equilateral, triangle.  See
Figure~\ref{isocelesfig}.  It is clear by construction that a half
rotation of the Bloch sphere about the $x$-axis fixes the point
$\ket{0}+\ket{1}$ and interchanges the other two vertices.  Thus it is
clear that $\Stab_\psi$ contains the order two element $XXX$.  While it
is a straightforward calculation\footnote{See~\cite{minorb1} for the
  linear system whose solutions are the local unitary stabilizer
  subalgebra.} $\;$ to find that $K_\psi=0$, it is not obvious from the
Bloch sphere picture that there is no continuous symmetry.  To
illustrate this point, consider the states
\begin{eqnarray*}
\ket{\phi}&=&\sum_\pi \pi\left[\ket{a}\ket{d}\ket{e}\right]\\
\ket{\phi'}&=&\sum_\pi \pi\left[\ket{a}\ket{d}\ket{e}\ket{f}\right]
\end{eqnarray*}
where $\ket{a}$, $\ket{b}$, $\ket{c}$ are as above and the remaining 1-qubit states are given by
\begin{eqnarray*}
\ket{d} &=& \frac{1}{2}\ket{0}+\frac{\sqrt{3}}{2}\ket{1}\\
\ket{e} &=& -\frac{1}{2}\ket{0}+\frac{\sqrt{3}}{2}\ket{1}\\
\ket{f} &=& \frac{1}{\sqrt{2}}(\ket{0}+i\ket{1})
\end{eqnarray*}
made by symmetrizing products of three vertices of an equilateral
triangle along a great circle in the case of $\ket{\phi}$, and adding a
fourth vertex spaced equally from the three vertices in $\ket{\phi}$ to
make $\ket{\phi'}$.  See Figure~\ref{2egfig}.  It turns out that
$\ket{\phi}$ is LU equivalent to the GHZ state, and therefore has two
dimensions of continuous symmetry with $K_\phi = \left\{t_1B^{(1)}+t_2B^{(2)}+t_3B^{(3)}\colon \sum t_k=0\right\}$, as well as being
a stabilizer state.  However, $\ket{\phi'}$ has no continuous symmetry,
but does have finite order stabilizing elements including the order 4 element
$e^{-\frac{2\pi i}{3}} \left( e^{\frac{2\pi i}{3}Y}\right)^{\otimes                                 
  4}$.  Thus the stabilizer of $\ket{\phi'}$ contains a cyclic group of
size 4, and one might guess that there are no additional elements, but
this is not obvious.  

We conclude with an example for which we can use Theorem~\ref{stabfinite} to say
what the stabilizer is.  Let $\ket{\tau}$ be the 6-qubit state
$$                                                                                                  
\ket{\tau}= \sum_\pi \pi\left[\ket{a}\ket{b}\ket{c}\ket{f}\ket{g}\ket{h}\right]                             
$$
where 
$$\ket{g}=\frac{1}{\sqrt{2}}(\ket{0}-\ket{1}), \hspace*{.2in} \ket{h}=\frac{1}{\sqrt{2}}(\ket{0}-i\ket{1})$$
made with the vertices of a regular octahedron.  By construction, the
stabilizer $\Stab_\tau$ contains the full 24 element octahedral group.
It follows from Theorem~\ref{stabfinite}, together with the known
classification of finite subgroups of $SO(3)$ (see, for example, \cite{weyl}), that this is the entire
stabilizer.

\begin{figure}
\begin{center}
\input{isocelessmall.pstex_t}
\fcaption{\label{isocelesfig}State with discrete symmetry but no continuous symmetry.}
\end{center}
\end{figure}

\begin{figure}
\begin{center}
\input{2egsmall.pstex_t}
\fcaption{\label{2egfig}Three-qubit state $\ket{\phi}$ has continuous symmetry, while
four-qubit state $\ket{\phi'}$ has no continuous symmetry.  Both
states have the one-third rotation about the $Y$ axis as a finite
order symmetry.}
\end{center}
\end{figure}

\begin{figure}
\begin{center}
\input{octosmall.pstex_t}
\fcaption{\label{octofig}Six-qubit state stabilized by the octohedral group.}
\end{center}
\end{figure}

\section{Conclusion}

We have completely classified symmetric states whose local unitary
stabilizers have positive dimension, and have shown that discrete
stabilizer groups are isomorphic to subgroups of $SO(3)$.  What remains
to be achieved to complete the LU classification is a difficult problem:
Given a discrete stabilizer subgroup, what states are stabilized by that
group?\\

A natural line of future investigation is to identify {\em families} of LU
classes that subdivide the infinite sets of LU classes that possess the
same discrete stabilizer, in the spirit of the classification given by
Bastin et al. in~\cite{Bastin} for the SLOCC case.\\

The Majorana constructions of the previous section provide a means of
finding states whose stabilizers contain desired finite order elements.
But it is not obvious what additional stabilizing elements will occur
beyond the symmetries of the geometric configurations on the Bloch
sphere of the 1-qubit states whose symmetrized product forms the given
state.  It would be useful to develop a stabilizer calculator that works
directly with the 1-qubit state configuration as input.\\

A natural generalization of symmetric group invariance is symmetric {\it
  subgroup} invariance.  An example that shows this is not a frivolous
exercise in abstraction is the state
$$ \ket{M_4} = \frac{1}{\sqrt{6}} [ \ket{0011} + \ket{1100} + \omega
  (\ket{1010} + \ket{0101}) + \omega^2 (\ket{1001} + \ket{0110}) $$
  where $\omega = \exp(2 \pi i/3)$, studied by Higuchi et
  al.~\cite{higuchi00,brierley07} and more recently by Gour and Wallach
  in~\cite{gour10}, maximizes average two-qubit
  bipartite entanglement, averaged over all partitions into 2-qubit
  subsystems.  This is an example of a maximally multipartite entangled
  (MME) state~\cite{facchi}.  This state is a CDC, and is stabilized by
  the four-element subgroup of the symmetric group whose elements are
  products of disjoint transpositions, but is not a symmetric state.
  This suggests that it may be fruitful to study states stabilized by
  subgroups of the symmetric group.\\

We expect that further analysis of states with discrete stabilizers will
reveal useful connections to other areas of quantum information.

\section{Acknowledgments}

This work was supported by National Science Foundation grant \#PHY-0903690.

\nonumsection{References}

\end{document}